\documentclass[aps,prd,preprint,groupedaddress,nofootinbib,superscriptaddress,floatfix,preprintnumbers,11pt]{revtex4-1}

\usepackage{graphicx}
\usepackage{amssymb}

\usepackage{lineno}
\usepackage[section]{placeins}

\usepackage{amsmath}

\usepackage{bm}
\usepackage{array}\allowdisplaybreaks[1]
\usepackage{braket}
\usepackage{slashed}
\usepackage{siunitx}
\usepackage{ulem,xpatch}
\usepackage{subfigure}
\usepackage{epsfig,color}
\usepackage{epstopdf}
\usepackage{enumerate}
\usepackage{appendix}
\usepackage{setspace}

\usepackage{comment}
\usepackage{soul}
\usepackage{xcolor}
\usepackage{hyperref}
\usepackage{xfrac}
\hypersetup{colorlinks, linkcolor = [rgb]{1,0,0}, citecolor = [rgb]{1,0,0}, urlcolor = [rgb]{1,0,0}}

\newcommand{\beq}{\begin{equation}}
\newcommand{\enq}{\end{equation}}
\newcommand{\beqa}{\begin{eqnarray}}
\newcommand{\beqast}{\begin{eqnarray*}}
\newcommand{\enqa}{\end{eqnarray}}
\newcommand{\enqast}{\end{eqnarray*}}
\newcommand{\req}[1]{(\ref{#1})}

\newcommand{\lb}{\label}

\newcommand{\bec}{\begin{center}}
\newcommand{\enc}{\end{center}}
\newcommand{\beqo}{\begin{quote}}
\newcommand{\enqo}{\end{quote}}

\newcommand{\al}{\alpha}

\newcommand{\ga}{\gamma}

\newcommand{\la}{\lambda}

\newcommand{\ta}{\tau}

\newcommand{\ps}{\psi}

\newcommand{\La}{\Lambda}

\newcommand{\Up}{\Upsilon}

\usepackage{soul}

\hypersetup{colorlinks, linkcolor = [rgb]{0, 0, 0.5}, citecolor = [rgb]{0,0.0,1.0}, urlcolor = [rgb]{0,0.0,0.5}}

\begin{document}

\preprint{JLAB-THY-22-3551}
\preprint{SLAC-PUB-17641}

\title{Towards a single scale-dependent Pomeron in holographic light-front QCD}

\newcommand*{\HEIDELBERG}{Institut f\"ur Theoretische Physik der Universit\"at, D-69120 Heidelberg, Germany}\affiliation{\HEIDELBERG}
\newcommand*{\COSTARICA}{Laboratorio de F\'isica Te\'orica y Computacional, Universidad de Costa Rica, 11501 San Jos\'e, Costa Rica}\affiliation{\COSTARICA}
\newcommand*{\SDU}{Key Laboratory of Particle Physics and Particle Irradiation (MOE), Institute of Frontier and Interdisciplinary Science, Shandong University, Qingdao, Shandong 266237, China}\affiliation{\SDU}
\newcommand*{\WM}{ Department of Physics, William \& Mary, Williamsburg, Virginia 23187, USA}\affiliation{\WM}
\newcommand*{\SLAC}{SLAC National Accelerator Laboratory, Stanford University, Stanford, CA 94309, USA}\affiliation{\SLAC}
\newcommand*{\JLAB}{Thomas Jefferson National Accelerator Facility, Newport News, VA 23606, USA}\affiliation{\JLAB}

\author{Hans~G\"unter~Dosch}%\email{dosch@thphys.uni-heidelberg.de}
\affiliation{\HEIDELBERG}
\author{Guy~F.~de~T\'eramond}\affiliation{\COSTARICA}
\author{Tianbo~Liu}\email{liutb@sdu.edu.cn}\affiliation{\SDU}
\author{Raza~Sabbir~Sufian}\affiliation{\WM}\affiliation{\JLAB}
\author{Stanley~J.~Brodsky}\affiliation{\SLAC}
\author{Alexandre~Deur}\affiliation{\JLAB}

\collaboration{HLFHS Collaboration}

\begin{abstract}

The Pomeron Regge trajectory underlies the dynamics dependence of hadronic total cross sections and diffractive reactions at high energies. The physics of the Pomeron is closely related to the gluon distribution function and the gluon gravitational form factor of the target hadron. In this article we examine the scale dependence of the nonperturbative gluon distribution in the nucleon and the pion which was derived in a previous article~\cite{deTeramond:2021lxc} in the framework of holographic light-front QCD and the Veneziano model. We argue that the QCD evolution of the gluon distribution function $g(x,\mu)$ to large $\mu^2$ leads to a single scale-dependent Pomeron. The resulting Pomeron trajectory $\alpha_P(t, \mu)$ not only depends on the momentum transfer squared $t$, but also on the physical scale $\mu$ of the amplitude, such as the virtuality $Q^2$ of the interacting photon in inclusive diffractive electroproduction. This  can explain not only the $Q^2$ evolution of the proton structure function $F_2(x,Q^2)$ at small $x$, but also the observed energy and $Q^2$ dependence of high energy diffractive processes involving virtual photons up to LHC energies.

\end{abstract}

\maketitle

\section{Introduction}

Despite the successful applications of perturbative quantum chromodynamics (pQCD) in describing hadronic physics at short distances, many complexities in the soft domain characterizing small momentum-transfer scattering processes at high energies remain unsolved. In practice, either phenomenological or model-dependent nonperturbative physics inputs are required  in order to predict high energy scattering and diffractive processes in the small longitudinal light-front momentum fraction $x$-domain.

In a previous article~\cite{deTeramond:2021lxc}, we  have studied the dynamics of gluons inside hadrons based on the gauge/gravity correspondence~\cite{Maldacena:1997re}, its light-front holographic mapping~\cite{Brodsky:2006uqa,deTeramond:2008ht,deTeramond:2018ecg}, and the generalized Veneziano model~\cite{Veneziano:1968yb,Ademollo:1969wd,Landshoff:1970ce}. Although an exact gravity dual to QCD  has yet to be discovered, the holographic light-front QCD (HLFQCD)  framework captures many  important nonperturbative features of QCD, including color confinement~\cite{Brodsky:2014yha}, chiral symmetry breaking~\cite{deTeramond:2021yyi} and the power-law falloff of the counting rules for hard scattering dynamics at large momentum transfer~\cite{Brodsky:1973kr, Matveev:1973ra, Lepage:1980fj}, which can be derived in the gauge/gravity correspondence from the warped geometry of anti-de Sitter (AdS) space~\cite{Polchinski:2001tt}. More recent insights, based on superconformal quantum mechanics~\cite{deAlfaro:1976je, Fubini:1984hf} and light-front quantization~\cite{Dirac:1949cp,Brodsky:1997de,Mannheim:2020rod} have  led to remarkable connections among  the spectroscopy of mesons, baryons and tetraquarks, as well as predicting a massless pion in the chiral limit~\cite{Brodsky:2013ar, deTeramond:2014asa, Dosch:2015nwa,  Brodsky:2020ajy}. This nonperturbative  color-confining formalism, {\it light-front holography}, incorporates the underlying conformality of QCD and describes an effective QCD coupling in the nonperturbative domain~\cite{Brodsky:2010ur, Deur:2016opc}.

In Ref.~\cite{deTeramond:2021lxc} we used the soft Pomeron trajectory as a key ingredient to compute the  gluonic gravitational form factor (gGFF) and the intrinsic gluon distributions in the pion and nucleon. The Pomeron trajectory was originally introduced~\cite{Chew:1961ev, Gribov:1961fr} to describe diffractive processes in terms of Regge theory.  The value of the Pomeron trajectory at zero momentum transfer, the Pomeron intercept, plays a special role: It determines the energy dependence of total cross sections at large energies~\cite{Donnachie:1992ny, Donnachie:2002en}. Since the work of~\cite{Low:1975sv, Nussinov:1975mw}, it  has been  generally accepted that gluon exchange is the essential dynamical mechanism  underlying diffractive processes~\cite{Donnachie:2002en};  this provides the connection between the soft Pomeron  trajectory and intrinsic gluon distributions, as discussed in Ref.~\cite{deTeramond:2021lxc}.  The  summation of gluon ladders derived from perturbative QCD introduces power-like energy dependence to the  diffractive cross sections~\cite{Fadin:1975cb, Kuraev:1977fs, Balitsky:1978ic}, which in turn, has led  to  the introduction of the  Balitsky-Fadin-Kuraev-Lipatov (BFKL)  ``hard Pomeron". Thus it has become conventional to assume the existence of two separate Pomerons~\cite{Donnachie:1998gm}, a soft and a hard one, with very different intercepts.

 By using the warped-space gauge/gravity framework for large-$N_C$ QCD-like theories, Brower, Polchinski, Strassler and Tan derived a simultaneous description of both the BFKL hard regime and the classic Regge soft domain~\cite{Brower:2006ea}. Their model is consistent with some salient general features which one would expect from the hard BFKL Pomeron at negative values of the momentum transfer $t$ and with a glueball spectrum at positive $t$. This model, however, did not solve the problem of the large difference of intercept values of the soft and the BFKL Pomeron. The basic idea introduced in Ref.~\cite{Brower:2006ea}, namely that the wave function of hadronic extended objects contains hard and soft components depending on the position of the object in the holographic coordinate $z$ in anti-de Sitter (AdS) space, was reexamined in Refs.~\cite{Hatta:2007he, Shuryak:2013sra, Iatrakis:2016rvj} for the Pomeron. Further studies of Pomeron exchange in the small-$x$ Regge domain using the gauge/gravity duality are described in Refs.~\cite{Cornalba:2008sp, Domokos:2009hm, Brower:2010wf, Costa:2012fw, Costa:2013uia}. On a phenomenological level, the question of a scale-dependent intercept was addressed in~\cite{Dosch:2015oha},  where  both the diffractive photoproduction data  up to LHC energies, as well as  the specific small-$x$ behavior of the  proton structure function $F_2(x,Q^2)$ have been described quantitatively by a scale-dependent Pomeron intercept.

Our approach to  the gluon distribution functions described in this article adds theoretical support to a single Pomeron with a scale-dependent intercept.  It  provides a natural way to compute intrinsic nonperturbative quantities at the hadronic scale,  which can then be evolved to higher scales using the renormalization group equations (RGE) of pQCD.  In this approach, the Pomeron intercept determines the small-$x$ behavior of the gluon distribution function of a hadron. Assuming that the relations between the Regge parameters and the gluon distribution  functions derived in~\cite{deTeramond:2021lxc} are also valid at higher scales, where RGE  contributions become important, we can  then determine the scale dependence of the Pomeron  intercept, which, in turn, can be used to explain the observed scale dependence of small-$x$ diffractive processes.  One can then address whether the soft Pomeron intercept evolved to high scales agrees with the larger intercept of the BFKL Pomeron and how   it can be related to diffractive processes at LHC energies.  We will examine these and other related questions in this article  within the HLFQCD framework and the generalized Veneziano model, with the aim of providing a unified model in which the soft Pomeron evolves to a BFKL Pomeron in high virtuality processes.

This article is organized as follows: In Sec. \ref{PI} we extract a scale-dependent Pomeron intercept from the nonperturbative gluon distribution obtained in~\cite{deTeramond:2021lxc} at the hadronic scale, which is then continued to higher scales using pQCD evolution equations~\cite{Gribov:1972ri, Altarelli:1977zs,Dokshitzer:1977sg}. In Sec. \ref{DP}, we compare this result with HERA measurements~\cite{Adloff:2001rw,Radescu:2013mka} of the proton's structure function. In Sec. \ref{SP} we present further evidence for a scale-dependent Pomeron intercept~\cite{Dosch:2015oha}, mainly based on photoproduction data at the LHC~\cite{LHCb:2014acg,ALICE:2014eof,ALICE:2018oyo,LHCb:2015log}. Sec. \ref{SC} contains a short summary and conclusions.

%%%%%%%%%%%%%%%%%
%
%%%%%%%%%%%%%%%%%

\section{\lb{PI}The Pomeron intercept and the gluon distribution in the hadron}

In previous articles~\cite{deTeramond:2018ecg,Sufian:2018cpj,Liu:2019vsn,deTeramond:2021lxc,Sufian:2020coz} we have derived parton distributions from the underlying hadronic form factors obtained in the HLFQCD framework within the constraints of the generalized Veneziano model~\cite{Veneziano:1968yb,Ademollo:1969wd,Landshoff:1970ce}. In this approach, the form factors are expressed in terms of the Euler Beta function. By comparing with the generalized Veneziano model including currents~\cite{Ademollo:1969wd,Landshoff:1970ce}, we deduced in~\cite{deTeramond:2021lxc} that the twist-$\ta$ Fock-state contribution to the    gravitational form factor of a hadron is given by 
\beq \label{A}
 A_\tau(t) = \frac{1}{N_\tau} B\big(\tau -1, 2 - \alpha_{P}(t)\big),
\enq
where  $\alpha_P(t)$ is  the soft Pomeron of Donnachie and Landshoff~\cite{Donnachie:1992ny}. It corresponds to a Regge trajectory,
\begin{align} \label{RP}
\alpha_{P}(t) = \alpha_{ P}(0) + \alpha'_{P} t, 
\end{align}
with intercept $\alpha_P(0) \simeq 1.08$  and  slope $\alpha’_P \simeq  0.25 \, {\rm GeV}^{-2}$~\cite{Zyla:2020zbs}. For integer twist $\tau$, the number of constituents of a given Fock state, the GFF~\req{A} is expressed as a product of $\tau -1 $ poles corresponding to the particles exchanged in the cross $t$-channel~\cite{deTeramond:2021lxc}. For large momentum transfer $-t = Q^2$ the expression~\req{A} reproduces the hard-scattering power behavior~\cite{Brodsky:1973kr, Matveev:1973ra}
\begin{align}
A_\tau(Q^2)  \sim \left( \frac{1}{Q^2} \right)^{\tau -1}.
\end{align}

The Euler Beta function $ B(u,v) = B(v,u)$ has the integral representation
\begin{align} \lb{B}
 B(u,v) = \frac{\Gamma(u) \Gamma(v)}{\Gamma(u+v)}
= \int_0^1 dy\, y^{u-1} (1-y)^{v-1},
 \end{align}
 with $\Re(u) > 0$ and $\Re(v) > 0$. The normalization factor  $N_\tau = B\big(\tau - 1, 2 - \alpha_P(0)\big)$  follows the convention given in~\cite{deTeramond:2021lxc}. Using Eq.~\req{B},  one can express the gravitational form factor as
\beq \label{Aw}
A_\tau(t) = \frac{1}{N_\tau}  \int_0^1 dx \, w'(x) w(x)^{1 - \alpha_P(t)} \big[1- w(x)\big]^{\tau-2}, \\
\enq
where the integrand can be identified with the generalized parton distribution at zero skewness $\xi$: $H(x,t)\equiv H(x,\xi=0,t)$ via
$ A_\tau(t) = \int_0^1 dx\,H_\tau(x,t)$. Its forward limit gives the twist-$\tau$ component of the gluon distribution function~\cite{deTeramond:2021lxc},
\beq \lb{g}
x\, g_\ta (x) = \frac{1}{N_\ta} w'(x) \, w(x)^{1-\al_P(0)}  [1 - w(x)]^{\ta-2}.
\enq
The   universal function $w(x)$ is independent of the twist $\tau$ and it satisfies the boundary conditions~\cite{deTeramond:2018ecg}:
\begin{align} \lb{w}
    w(0) = 0, 
    \quad 
  w(1) = 0, 
    \quad
    w'(1) = 0,
    \quad
    w'(x) > 0~~{\rm for}~~0\leq x < 1, 
\end{align}
which largely determines its behavior. Physical constraints  can be imposed on $w(x)$ at small and large $x$ : At $x \to 0$, $w(x) \sim x$ from Regge theory~\cite{Regge:1959mz}, and  at $x \to 1$, one can apply the inclusive-exclusive counting rule~\cite{Drell:1969km,Brodsky:1979qm},
$g_\tau(x) \sim (1-x)^{2 \tau - 3}$, which fixes the additional condition $w'(1) = 0 $.
A convenient parametrization of $w(x)$, which fulfills all  of these constraints \req{w} is 
\begin{align} \lb{wxb}
w(x) = x^{1-x} \, e^{-b(1 -x)^2},
\end{align}
where $b$ is a parameter determined from phenomenology,  which is  fixed by the first moment of the nucleon unpolarized valence quark distribution. The value $b=0.48\pm0.04$ gives a good description of the quark and gluon distributions~\cite{Liu:2019vsn,deTeramond:2021lxc} for nucleons as well as for the pion.

The gluon distribution in a hadron is the sum of contributions from all Fock states which contain a gluon component,
\begin{align}\label{xg}
    x g(x) = \sum_{\tau} c_\tau x g_\tau(x),
\end{align}
where the $c_\tau$'s are expansion coefficients of the corresponding Fock states. In practice, one has to truncate the expansion at some value of $\tau$ for phenomenological studies. It has been found in our recent work, that the leading Fock components containing one dynamical gluon, $|uudg\rangle$ for the proton, with $c_{\tau=4} = 0.225\pm 0.014$, and $|u\bar{d} g\rangle$ for the pion, with $c_{\tau=3} = 0.429 \pm 0.007$, provide a satisfactory description of the gluon distributions in the proton and pion~\cite{deTeramond:2021lxc}, using the same universal function $w(x)$ as for the proton and pion quark distributions. The coefficients $c_\tau$ are determined from the momentum sum rule using the previous results given in Refs.~\cite{Liu:2019vsn, deTeramond:2021lxc}.  A similar Fock-state configuration including one dynamical gluon for the pion is described in~\cite{Lan:2021wok}.

The parton distribution functions, including the gluon distribution, are not direct observables and are scale and  renormalization-scheme dependent. The gluon distribution given in Eq.~\eqref{g} should be understood as being determined at an initial nonperturbative scale $\mu_0$. We choose the value $\mu_0=1.057 \pm  0.15\,\rm GeV$  which is determined by matching the strong coupling $\alpha_s(Q^2)$ between its perturbative expansion in the high energy region and the HLFQCD expression of the effective strong coupling in the low energy region~\cite{Deur:2016opc, Deur:2016cxb}. This value for $\mu_0$ is consistent with the Veneziano model, which is a nonperturbative model  applicable at the hadronic scale. Accordingly, the gluon distribution corresponds to ``intrinsic gluons'', which exist for a long time scale in the wave function of the  hadronic eigenstate. It has a different physical origin from the ``extrinsic gluons" originating in the pQCD parton splitting process, such as triggered by an external hard collision.

\begin{figure}[htp]
\bec
\includegraphics[width=0.5\textwidth]{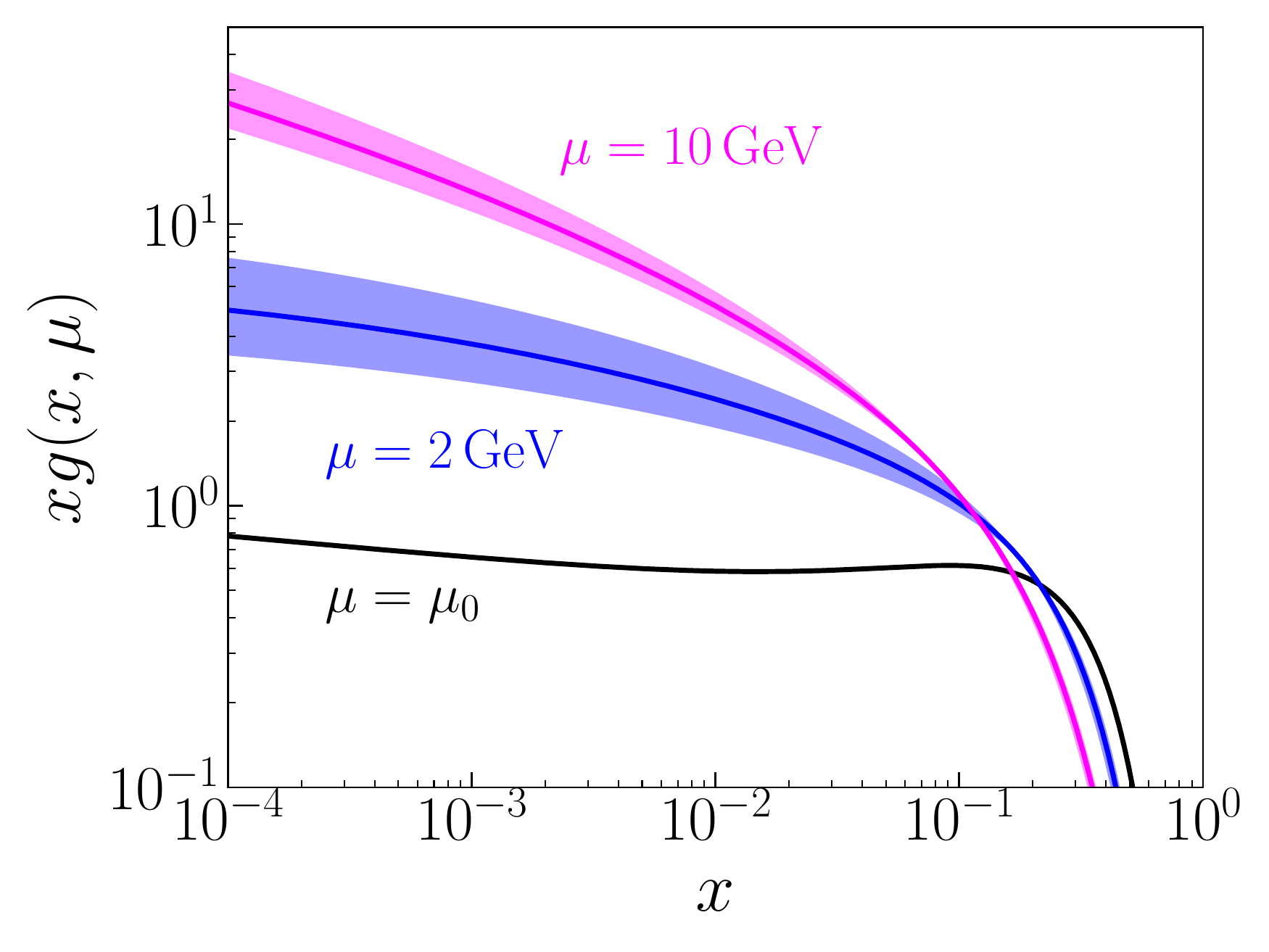}
\enc
\caption{Gluon distribution function in the nucleon $x g(x)$ at $\mu=2\,\rm ~ GeV$ (blue), and $\mu =10\,\rm ~GeV$ (magenta) using DGLAP evolution from the initial hadronic scale $\mu_0$ (black).  The uncertainty band stems from the initial scale uncertainty $\mu_0=1.06\pm0.15\,\rm GeV$.~\label{fsd}}
\end{figure}

Thus, in order to evaluate the full gluon distribution, which includes both intrinsic and extrinsic contributions, one needs to take into account the gluons generated from splitting processes, such as $q\to qg$ and $g\to gg$, in the Dokshitzer–Gribov–Lipatov–Altarelli–Parisi (DGLAP) evolution equations of pQCD~\cite{Gribov:1972ri,Altarelli:1977zs,Dokshitzer:1977sg}. Following this standard procedure, the parton distributions at a  high scale $\mu$ are determined by the input distributions at a lower initial scale $\mu_0$, as long as QCD perturbation theory remains valid  within the chosen range of scale evolution.  Since QCD is flavor-blind, the gluon distribution only mixes with the flavor-singlet combination of the quark distributions. At the input scale $\mu_0$ around $1\,\rm GeV$,  we can neglect contributions from intrinsic heavy quarks, and thus the flavor-singlet quark distribution at the initial scale corresponds to light quarks. For the proton, we take the quark distributions obtained in Refs.~\cite{deTeramond:2018ecg, Liu:2019vsn} and for the pion, the flavor-singlet twist-2 and twist-4 Fock state contributions obtained in Ref.~\cite{Brodsky:2014yha}, although the flavor separation is not uniquely determined in this case~\cite{deTeramond:2021lxc}. The numerical results are computed with the {\sc hoppet} toolkit~\cite{Salam:2008qg} at next-to-next-to-leading order, with the dominant uncertainty arising from the choice of the initial scale $\mu_0 = 1.06\pm 0.15\,\rm GeV$. As shown in Ref.~\cite{deTeramond:2021lxc}, the result agrees well with the full gluon distribution extracted from global analyses~\cite{Ball:2017nwa, Hou:2019efy, Bailey:2020ooq} for the proton and~\cite{Novikov:2020snp,Cao:2021aci} for the pion.  In Fig.~\ref{fsd}, we show the leading-twist $\tau= 4$  intrinsic gluon distribution in the proton at the hadronic scale $\mu_0$, obtained from \req{g}, together with the evolved predictions at $\mu =2\,\rm GeV$ and $10\,\rm GeV$.

%%%%%%%%%%%%%%%%%%%%%%%%%%%%%
%%%%%%%%%%%%%%%%%%%%%%%%%%%%

Since Eq.~\eqref{g} describes the gluon distribution in the full range of $x$, it is plausible to assume that it will maintain its functional form when continued to higher scales, including its scale dependence. Thus we write
\beq \lb{gmu}
x\, g(x,\mu) = \sum_\ta  \frac{1}{N_\ta(\mu)} c_{\ta}(\mu) w'(x,\mu) [1 - w(x,\mu)]^{\ta(\mu)-2}\, w(x,\mu)^{1-\al_P(0,\mu)},
\enq
where the scale-dependence  in $\mu$ can arise from several sources: the normalization $N_\tau(\mu)$, the Fock expansion coefficient $ c_{\ta}(\mu)$, the rescaling function $w(x,\mu)$, a scale-dependent effective twist $\ta(\mu)$, and a scale-dependent Pomeron intercept $\al_P(0,\mu)$. At first sight,  it appears intractable  to disentangle the   scale dependencies arising from  these different origins, but fortunately, the small-$x$  behavior is determined exclusively  by the intercept $\al_P(0,\mu)$, and therefore the latter can be extracted unambiguously.

%%%%%%%%%%%%%%%%%%%% 
\begin{figure}
\begin{center}
\includegraphics[width=10cm]{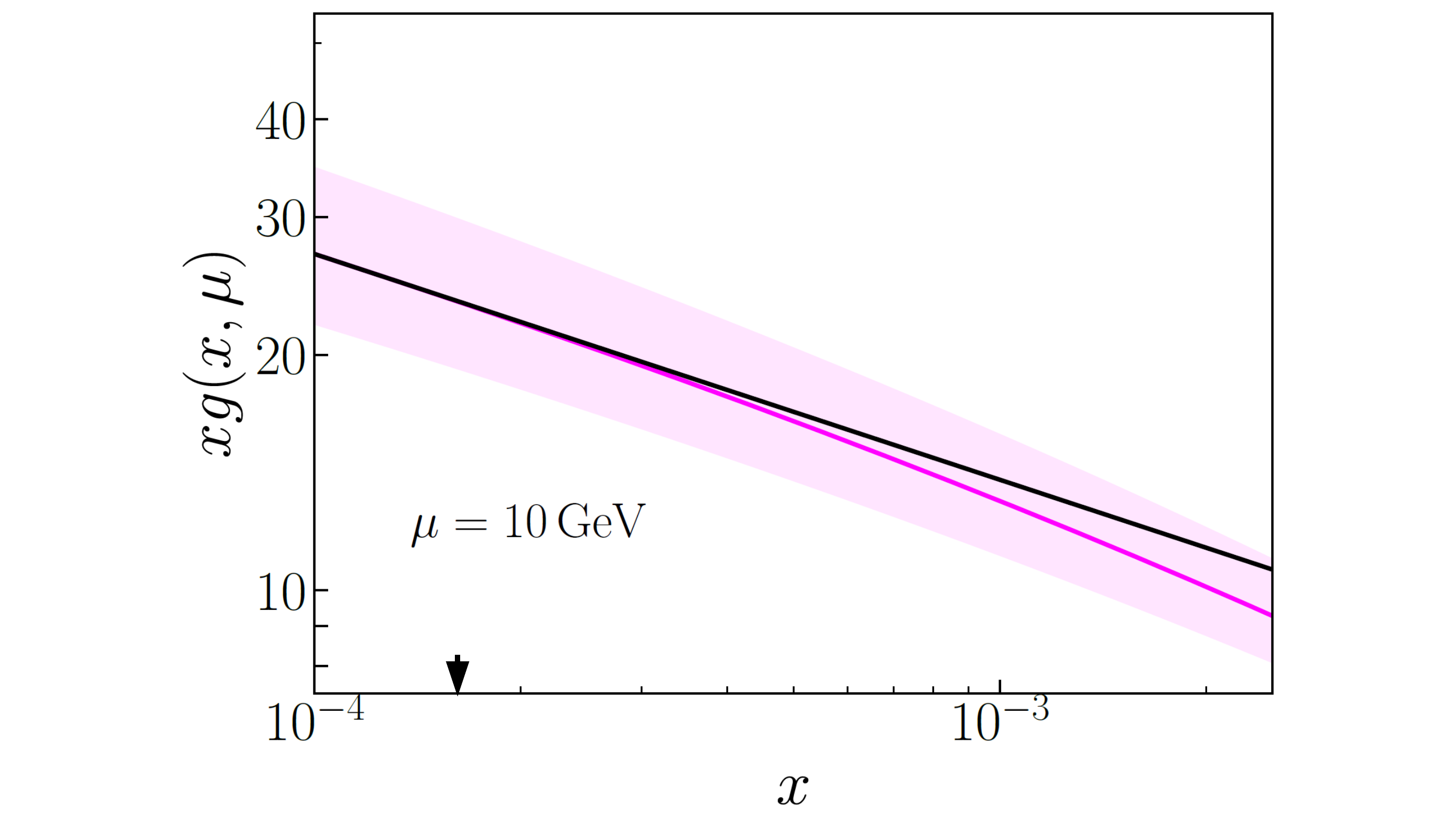}
\end{center}
\vspace{0cm}
\caption{\lb{f1} The gluon density $x\,  g(x, \mu)$ in the proton at the scale $\mu = 10$ GeV. Magenta: numerical result from the DGLAP evolution of the intrinsic gluon distribution in \cite{deTeramond:2021lxc}, black: leading terms of  its Laurent expansion  in the in the interval $10^{-4} \leq x \leq 1.6\times 10^{-4}$ (indicated by an arrow).  From the functional form \req{laurent} it follows that $ 1 - \alpha_P(0, \mu)$ is the slope of the linear approximation in the log-log plot. The uncertainty band corresponds to the initial scale uncertainty.}
\end{figure}
%%%%%%%%%%%%%%%%%%%

Making a Laurent expansion of $\log\big(x \,g(x,\mu)\big)$, Eq. \req{gmu}, in powers of $ 1/\log x$,
\beq \lb{laurent}
\log \big(x \,g(x,\mu)\big) = \big(1-\al_P(0,\mu)\big) \log x +  B(\mu) + O(1/\log x),
\enq 
we obtain the  Regge intercept from the expression $1-\al_P(0,\mu)$, the factor of the leading $\log x$ term in \req{laurent}. The next term $B(\mu)$ does not enter explicitly into our analysis. Therefore, the normalization of the gluon component of the gravitational form factor does not affect our result for $1-\al_P(0,\mu)$. The independence of our analysis on $B(\mu)$ also implies the independence of our results on the specific form of the universal function $w(x)$, since  $ w(x) \to x$ in the limit $x\to 0$.

 As a specific example of the procedure used to extract the scale dependence or the Pomeron Regge intercept, the numerical results for $\log\big(x\, g(x,\mu)\big)$ obtained  in Ref.~\cite{deTeramond:2021lxc} from the pQCD evolution of the intrinsic gluon distribution is compared in Fig.~\ref{f1}  with the first two terms of its small-$x$ Laurent expansion \req{laurent}  at the value $\mu=10$ GeV. The value of the Pomeron effective intercept at a given scale $\mu$, namely $ 1 - \alpha_P(0, \mu)$, is the slope of the linear approximation in the log-log plot in Fig~\ref{f1}.  The value of the second term in~\req{laurent}, $B(\mu)$, is also determined numerically from the perturbative evolution of gluon distributions. Its actual value has, however, no relevance for the present analysis. In Table~\ref{t1} we specify the values of $1-\alpha_P(0,\mu)$ at different evolution scales obtained from the expansion (10) in the range $0.0001 \leq x \leq 0.00016$. We also list in Table~\ref{t1}  the gluon component of the proton and pion gravitational form factors, $A^g(t=0,\mu)$,
\beq
A^g(0,\mu) = \int_0^1 dx \,x g(x,\mu),
\enq
for the leading twist $\tau = 4$ and $\tau = 3$ respectively, which is the momentum fraction carried by the gluon at the scale $\mu$.  The uncertainty in the choice of the starting point of the evolution mainly affects the  $x$-independent term $B(\mu)$ in the expansion \req{laurent}; this has minimal affect on the leading term linear in $\log x$ and is thus not relevant for the present analysis.

%%%%%%%%%%%%%%%%%%%
\begin{table}[htp]
\caption{\lb{t1} The values of the gluon component of the proton and pion gravitational form factors,  $A^g_p(0, \mu)$ and $A^g_\pi(0, \mu)$, and the effective Pomeron intercept $1-\al_P(0,\mu)$ are indicated for different scales. The first row corresponds to the initial hadronic scale  $\mu_0=1.06\pm0.15\,\rm GeV$.}
\bec
\begin{tabular}{c c c c}
\hline \hline \vspace{2pt}
$\mu$  (GeV) & $~~A^g_p(0, \mu)$ & $A^g_\pi(0,\mu)$ & \quad $1-\al_P(0,\mu)$\\
\hline   \vspace{2pt}
$\mu_0$  & $ 0.225\pm0.014$ & $~~~0.429\pm0.007$ &-0.08\\
2        &$ 0.318\pm 0.020$& $~~~0.464\pm0.009$ &  ~~ $ -0.097\pm0.018$\\
5        &$ 0.372\pm 0.015$& $~~~0.481\pm0.006$ & ~ $ -0.234\pm0.018$\\
10     &$ 0.390\pm 0.012$& $~~~0.482\pm0.005$&  ~ $ -0.292\pm0.017$\\
20     &$ 0.402\pm 0.010$& $~~~0.482\pm0.004$&  ~ $ -0.336\pm0.016$\\
50   &$ 0.413\pm 0.008$& $~~~0.482\pm0.003$&  ~ $ -0.381\pm0.015$ \\
100    &$ 0.419\pm 0.007$& $~~~0.482\pm0.002$&  ~ $ -0.407\pm0.015$ \\
\hline \hline
\end{tabular}
\enc
\end{table}
%%%%%%%%%%%%%%%%%%%%%%%%%%%%%%%%%%%%

\section{\lb{DP} Scale dependence of diffractive processes}

Diffractive processes, in which the scattered particle keeps its quantum numbers, are dynamically described in Regge theory~\cite{Collins:1977jy} by the exchange of particles with the quantum numbers of the vacuum, {\it i.e.,} by the Pomeron with trajectory $\alpha_P(t)$. According to the optical theorem, the total inclusive cross section $\sigma_{\rm tot}$ of the reaction is proportional to the imaginary part of the elastic forward scattering amplitude, and therefore its high energy dependence is determined by the value of the Pomeron trajectory at zero momentum transfer $\al_P(0)$,
\beq
\sigma_{\rm tot}(s) \sim s^{\alpha_P(0)-1},
\enq
where $s$ is the center of mass (CM) energy square of the colliding particles. The high energy behavior of inclusive hadron cross sections is well described by a hypercritical Pomeron~\cite{Donnachie:1992ny}, with intercept $\al_P(0) = 1.08$. We note that the resulting high energy behavior is ultimately incompatible with general principles of quantum field theory~\cite{Froissart:1961ux}, and therefore unitarity corrections (Regge cuts) are necessary in order to modify the behavior at extremely high energies.

The lepton-hadron scattering process  at the lowest order of the electromagnetic fine structure constant $\al_{\rm em} \approx 1/137$ can be viewed as the interaction of a virtual photon $\gamma^*$ and the hadron $h$, as illustrated in Fig. \ref{fs}. In  such processes, the amplitude depends on the photon virtuality, $Q^2 = - p_{\gamma^*}^2$, in addition to the total energy squared $s$ of the photon-hadron system. This allows one to introduce the dimensionless quantity $x_{bj}$, the Bjorken variable~\cite{Bjorken:1968dy}, 
\begin{align} \lb{xbj}
 { x_{bj}} = \frac{Q^2}{W^2 + Q^2 - M_h^2}, 
\end{align}
where $s= (p_{\ga^*}+p_h)^2  = W^2$  and $M_h^2 = p_h^2$. In deep inelastic scattering (DIS), one identifies $x_{bj}$ with the longitudinal light-front variable $x= \frac{k^+}{P^+}$ at leading twist. The $x$ dependence of electromagnetic scattering processes is a principal source of information about the inner structure and dynamics of hadrons.

%%%%%%%%%%%%%%
\begin{figure}
\bec
\includegraphics*[width=7cm]{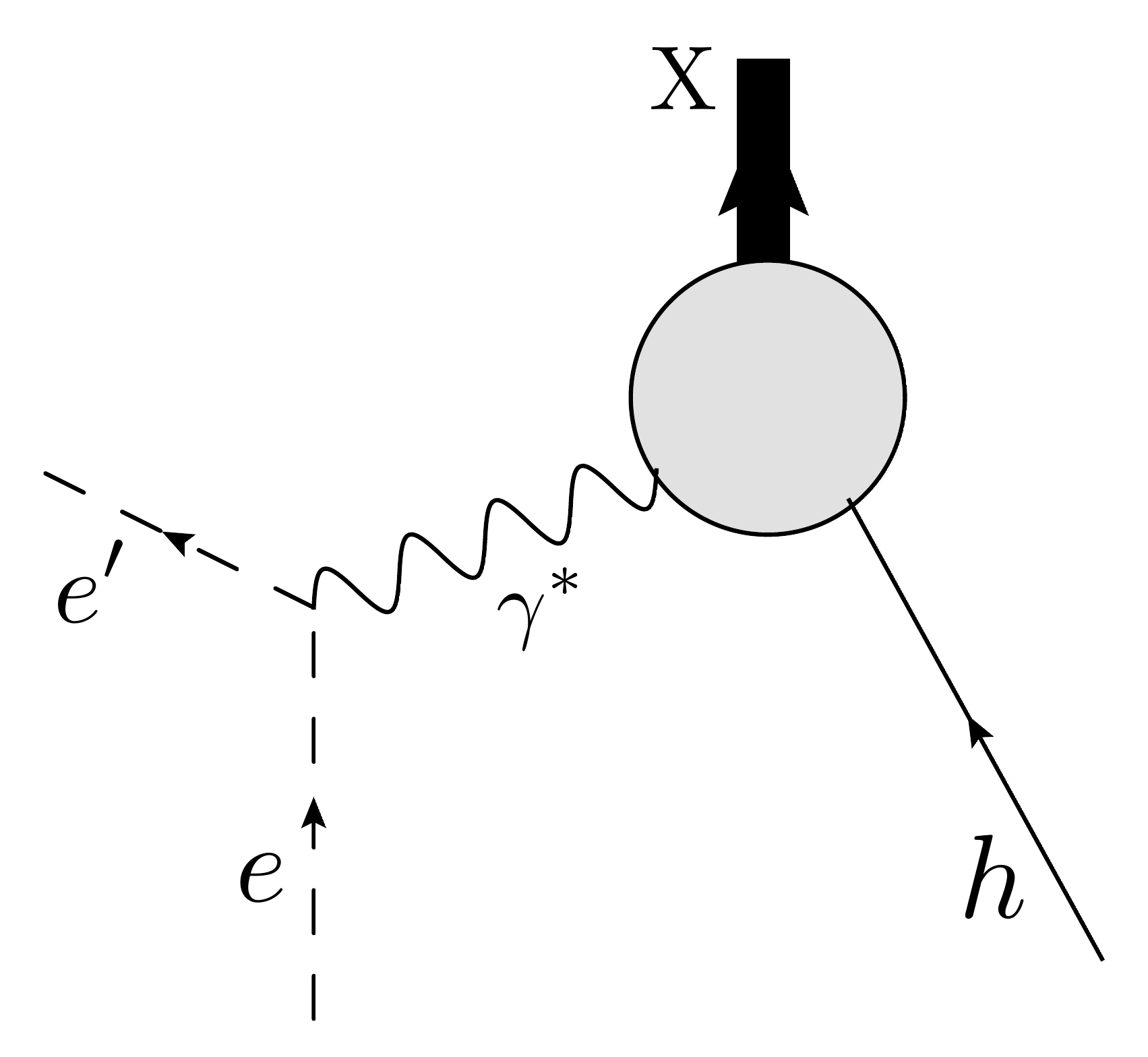}
\enc
\caption{\lb{fs}The reaction $e+h \to e'+ X$ can be viewed as the total inclusive cross section of a hadron $h$ and  {an off-shell} photon with virtuality $Q^2 =-(p_{e'}-p_e)^2 = -p_{\ga^*}^2$.}
\end{figure}
%%%%%%%%%%%%%

It is conventional to extract from the electron-hadron scattering cross section the {\it structure functions} of the hadron. One of them, $F_2(x,Q^2)$, can be directly related to the total  $\gamma^* h$ transverse cross section
\beq
\sigma^{\ga^* h}_{T} \approx \frac{4 \pi^2 \al_{\rm em}}{Q^2} F_2(x,Q^2).
\enq
 In the region of $3.5 \leq Q^2 <150\,\rm GeV^2$ and $x \leq 0.001$, the structure function can be fitted by a single power~\cite{Adloff:2001rw,Radescu:2013mka},
\beq \lb{la} 
F_2(x,Q^2) = c ~ x^{-\la_P(Q^2)},
\enq
where 
\beq 
\lb{hera} \la_P(Q^2) = 0.0481 \log \left(\frac{Q^2}{\Lambda^2} \right),  
\enq 
with  $\Lambda = 0.292\,{\rm GeV}$. In the high-energy domain $s = W^2 \gg Q^2 = -t$, one has \req{xbj} $x \sim Q^2/s$, therefore  the relation~\req{la}  corresponds to the energy behavior of the total cross section
\beq \lb{sv}
 \sigma^{\ga^* h}_T \sim s^{\la_P(Q^2)},
\enq
where the power of $s$ is definitely much greater than the value  expected from a simple Regge picture: From Eq.~\req{hera} it follows, for example, that  $\la_P(3.5 \,\rm GeV^2) \approx 0.18$  and  $\la_P(150\, {\rm GeV^2}) \approx 0.33$.

In QCD, the simple two-gluon exchange between two color neutral hadrons yields a constant cross section.  In contrast, the exchange of a gluon ladder  leads to  short-distance power behavior. At the lowest order in the strong coupling $\al_s$ one  obtains~\cite{Fadin:1975cb,Kuraev:1977fs,Balitsky:1978ic}
\beq
\al_{P}(0)-1 = \frac{12 \al_s}{\pi} \log 2 ~ { \simeq 2.65 \, \al_s},
\enq
but the next-to-leading order corrections are very large~\cite{Ciafaloni:1998gs,Fadin:1998py}.  One can therefore apparently conclude  that the gluon-ladder approximation  in pQCD yields a Pomeron trajectory with an intercept much greater than the one obtained  from hadron phenomenology. Motivated by this result and the $Q^2$ dependence of electromagnetic diffractive processes, Donnachie and Landshoff~\cite{Donnachie:1998gm} introduced a  perturbative BFKL-Pomeron with an intercept $\al_{\mbox{\rm \tiny BFKL}}(0)=1.42$, in addition to the ``nonperturbative''  Pomeron with intercept $\alpha_P(0) \approx 1.08$.  By following this procedure, the full structure function at small $x$,  its specific  heavy flavor contributions,  as well as the electroproduction of vector mesons, could be well described~\cite{Donnachie:2002en}.

As  noted above, the energy dependence of the measured total virtual-photon-proton cross section increases with $Q^2$ as $s^{\la_P(Q^2)}$. This increase could be directly explained by a  $Q^2$-dependent Pomeron intercept,
\beq \lb{lamu} 
\al_P(0,Q)-1 =\la_P(Q^2).
\enq 
The  analysis of the gluon distribution obtained in Ref.~\cite{deTeramond:2021lxc} directly supports such a concept. As emphasized above, the perturbative evolution of the intrinsic gluon distribution,  together with  the flavor-singlet quark distribution at the initial scale,  provides a good description  of the full gluon distribution without any additional input. There is no sign of an additional independent contribution from perturbative QCD. This makes the  conventional assumption of two Pomerons, a soft one due to the intrinsic gluon distribution and a hard one due to perturbative contributions, less convincing. We therefore  postulate that there is only  a single Pomeron $\al_P(t,\mu)$, which manifests itself at hadronic scales  $\mu \simeq  1$ GeV   as the soft Pomeron with an intercept at $t=0$ about 1.08, but  that the intercept is shifted by short-distance QCD interactions to larger values. Such a scale-dependent Pomeron intercept is fully compatible with the fundamental principles of Regge theory~\cite{Dosch:2015oha}.

%%%%%%%%%%%%%%%%%%%%%%%
\begin{figure}
\centering
\includegraphics[width=10cm]{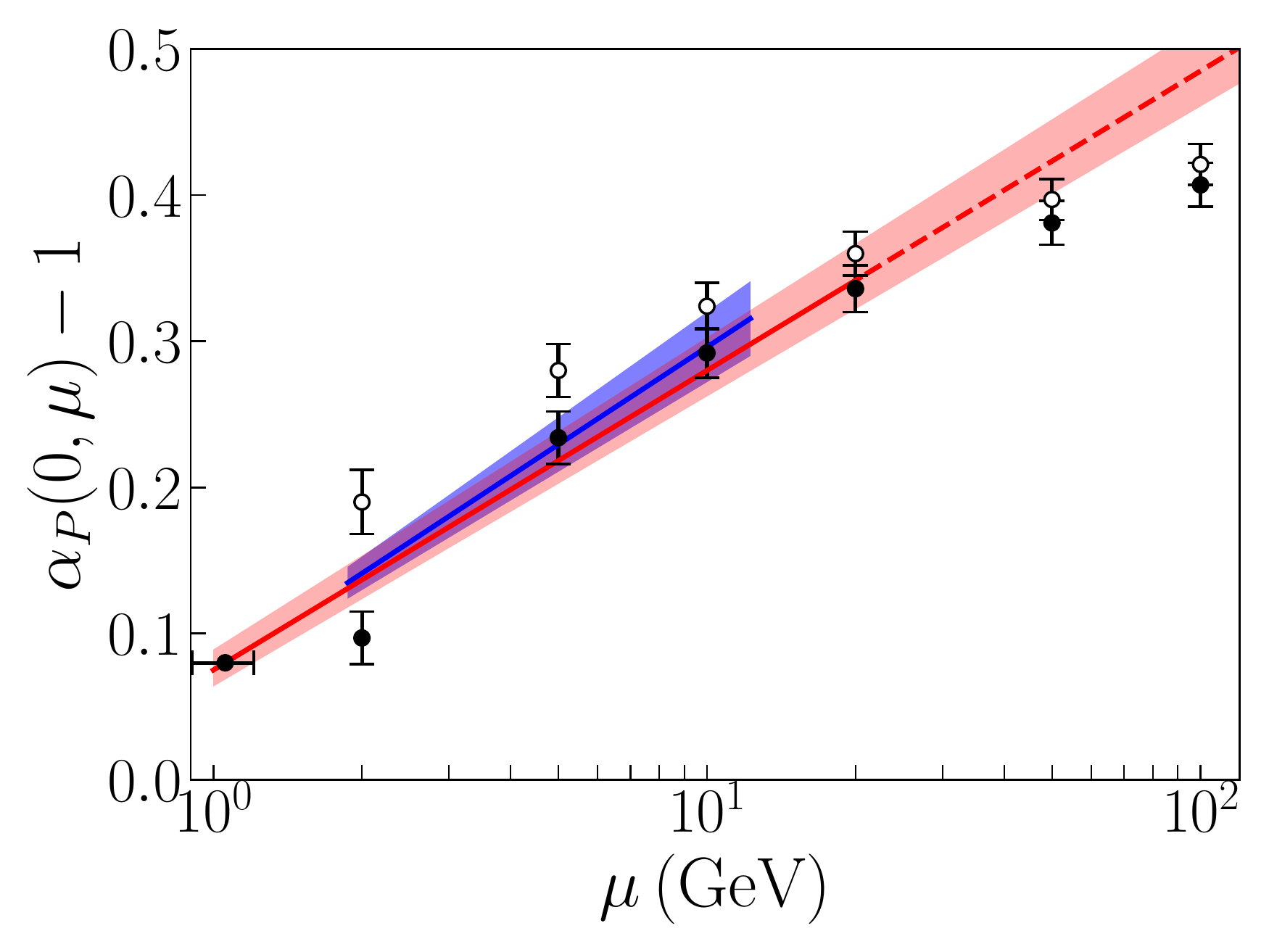}
\caption{\lb{mudep}  The Pomeron intercept $\al_P(0,\mu)-1$ extracted from the gluon distribution functions of the proton and the pion compared with the measured proton structure function. Black circles: values extracted with the Laurent expansion \req{laurent} from Eq.~\req{gmu} at $\mu$ = 1.06, 2, 5, 10, 20, 50, and 100 GeV for the proton and empty circles for the pion; red curve: linear fit from the points between $\mu =1.06$ and $\mu=  20$ GeV (the dashed part is the extrapolation to $\mu=100$  GeV); blue curve : $\la_P(Q^2)$ from the fit \req{hera} of the proton structure function $F_2(x,Q^2)$~\cite{Adloff:2001rw} with the subtraction point $\La_Q$ shifted to 0.461 GeV, such that $\la_P(1.06^2) = 0.08$.  The blue band represents the experimental uncertainty.}
\end{figure}
%%%%%%%%%%%%%%%%%%%%%%%%%% 

 In Fig.~\ref{mudep} we compare the values of $\al_P(0,\mu)-1 $ deduced  from the fit~\req{laurent} to the small $x$ behaviour of the gluon distribution functions of the proton and the pion at $\mu=1.06, 2, 5, 10, 20, 50$ and $100\,\rm GeV$ with the measured proton structure function: The full circles refer to the proton, the open ones to the pion. A logarithmic fit for the proton values  at $\mu=1.06, 3, 5, 10$ and $20\,\rm GeV$ leads to 
\beq \lb{almu}
\al_P(0,\mu)-1 = 0.08 + C \log \left(\frac{\mu}{\mu_0}\right),
\enq 
with $C=0.089\pm0.003$  for $\mu_0=1.06\pm0.15\,\rm GeV$.  The result is displayed as a red solid line in Fig.~\ref{mudep}; the dashed red line is the extrapolation up to $\mu = 100$ GeV. 
 
Measurements~\cite{H1:2013okq,ZEUS:2002wfj} of  the proton structure function $F_2(x,Q^2)$ cover the $Q$ range up to ${ 12.5}\,\rm GeV$.
The blue curve in Fig.~\ref{mudep} is the fit \req{hera} to the proton structure function $F_2(x,Q^2)$  given in Ref.~\cite{Adloff:2001rw}.  In order to compare the result of this article with the parametrization~\cite{Adloff:2001rw}  of the $x$-dependence  of the structure function $F_2(x,Q^2)$, Eq.~\req{hera}, one has to relate the renormalization scale $\mu$ in the gluon distribution function $x\,g(x,\mu)$  to  the photon virtuality $Q$. This is done by introducing a ``hadronic scale" $Q_0$ for the photon virtuality such that  $\la(Q_0^2)=1.08$.  This leads to the relation $Q=0.633\, \mu$. It should be noted, that the logarithmic slope of $\la(Q^2)$, the measure for the scale dependence of the Pomeron intercept, is not affected by this choice.

As can be observed in Fig.~\ref{mudep}, there is  good agreement between the energy dependence of the intercept derived from the evolution of the intrinsic gluon distribution of the proton and the results obtained from the total proton structure function $F_2(x, Q^2)$.  This agreement is not unexpected, since for small  $x$-values and  $Q$-values above the hadronic scale, the structure function is dominated by the gluon distribution.  Yet, since the theoretical value for the gluon distribution at high scales was obtained by the evolution of the intrinsic gluon distribution  from the hadronic scale, this result strongly supports the assumption that there are  not two Pomerons, but only one, where the values of its intercept (and presumably of its slope) are modified at large photon virtuality by pQCD interactions.

One would expect that the intercept derived from the gluon distribution function of the pion has the same value as that derived from the gluon distribution function of the nucleon. We  will discuss this in Sec.~\ref{SP}.

%%%%%%%%%%%%%%%%%%%%%%%%

\section{\lb{SP}Further arguments for a single Pomeron}

In this section we  review earlier arguments for a single, but scale-dependent, Pomeron trajectory.  We note that a Pomeron with scale-dependent parameters has been discussed qualitatively on the basis of holographic models in Refs.~\cite{Brower:2006ea,Hatta:2007he}. More recently, a quantitative investigation has been performed in Ref.~\cite{Dosch:2015oha}. 

%%%%%%%%%%%%%%%%%%%%%%%
\begin{figure}
\bec
\includegraphics*[width=7cm]{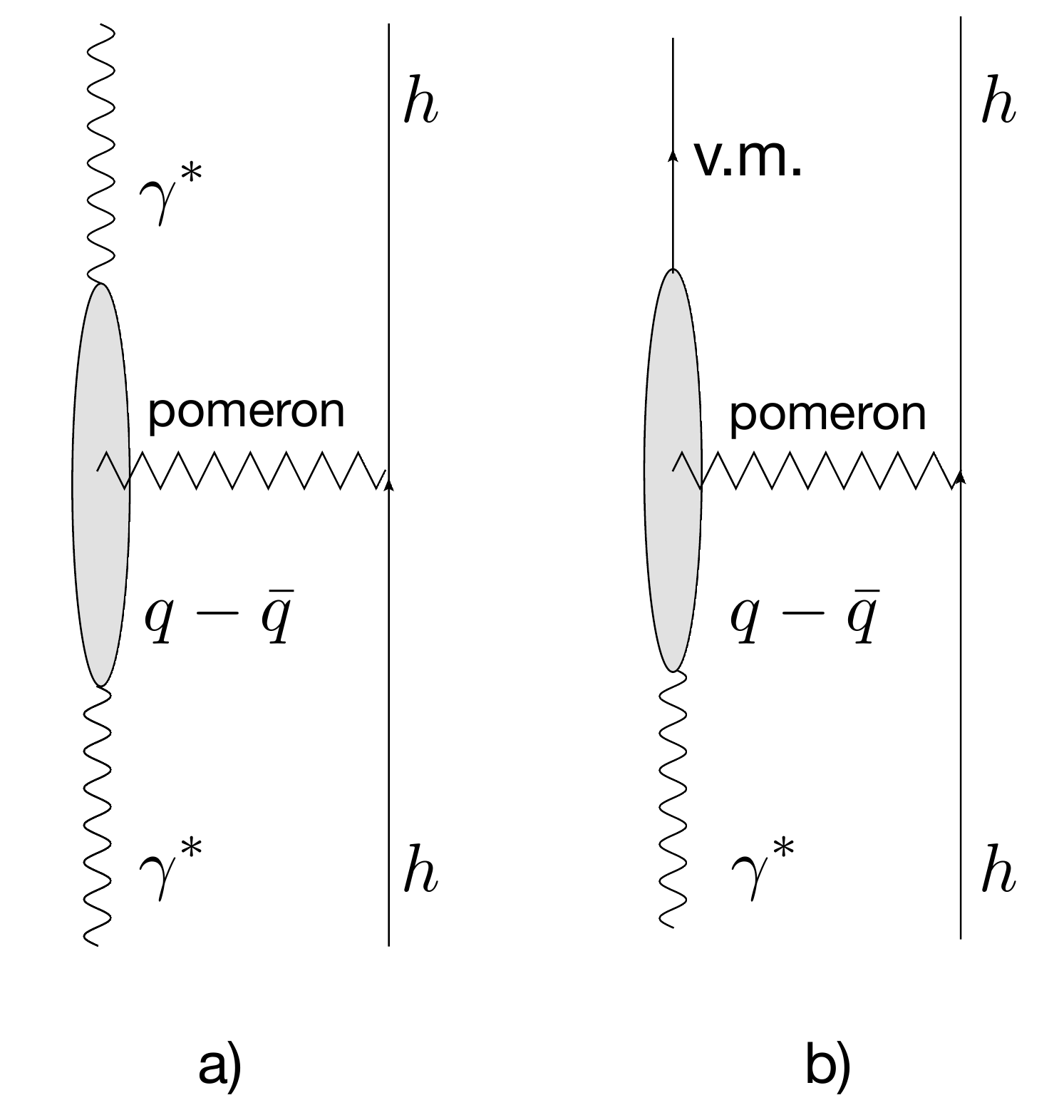}
\enc
\caption{\lb{vm} Diffractive $\ga^*$-hadron scattering (a) and diffractive electroproduction of vector mesons (b). The photon dissociates into a quark-antiquark pair; the lifetime of this pair is long compared to the interaction time. The scale is set by the  decay width of the $q-\bar q$ pair.}
\end{figure}
%%%%%%%%%%%%%%%%%%%

The  scale dependence  of diffractive cross sections can also be observed in diffractive production of vector mesons.\footnote{In  photoproduction processes the  minimum  value of momentum transfer is given by $t\leq t_{m} = -\frac{2\,M_V^4}{s} \,\left(1+ O\left( \frac{M_V^2}{s}\right)\right)$ where $M_V$ is the mass of the produced meson and $\sqrt{s}$ the  center of mass energy. For high energy processes the  value of $t_m$ is very close to zero.  Henceforth we always refer to  $\al_P(t_m,\mu)$  when discussing the intercept of vector meson production processes.} In such processes, the photon  dissociates into a quark-antiquark pair, which interacts with the hadron by Pomeron exchange, as illustrated in Fig.~\ref{vm} (b). The dissociation time at high energies can be shown~\cite{Ioffe:1969kf} to be much larger than the interaction time. The scale dependence can enter here through both the photon virtuality and the quark-pair mass.  In the case of photoproduction of $\rho$ mesons, the intercept is identical to the hadronic one: $\al_P(0) \approx 1.08$.  The intercept increases to about $1.2$ at $Q^2=30\,\rm GeV^2$; for the photoproduction of $J/\psi$ mesons, it is around $1.17$ and for $\Upsilon$ mesons, it is around 1.25~\cite{H1:2013okq,ZEUS:2002wfj}. Another possibility  for observing the scale dependence of diffractive processes is  to identify the  heavy-flavor contributions to the structure function $F_2(x, Q^2)$, which increase faster with energy than 
the light-flavor contributions;  see, {\it e.g.}, Ref.~\cite{Donnachie:2001wt}.

As mentioned above, in the usual Regge approach the scale dependence of diffractive processes is explained by the presence of two Pomerons with different intercepts~\cite{Donnachie:1998gm}: one at 1.08 and one around 1.42. If this is the case, the contribution of the Pomeron with the larger intercept  would become more dominant with increasing energy; therefore the slope of the energy dependence  would increase with increasing energy  leading to a convex dependence of the cross section on energy. Such a behavior could not be excluded by  measurements at the DESY storage ring~\cite{H1:2013okq,ZEUS:2002wfj}; however, the photoproduction data at the LHC tends to exclude this behavior,  favoring a global description by a single Pomeron with a scale-dependent intercept.

The integrated cross  section for the reaction $\ga + p \to J/\ps + p$  has been measured for total center of mass  energies $W$ up to 7 TeV. Within the range covered by the HERA data~\cite{H1:2013okq,ZEUS:2002wfj} there is no indication of a convex energy dependence and a straight  linear fit corresponding to an intercept of $\al_P(0,\mu) -1= 0.17$ can describe the data well.  Including the new photoproduction data at the LHC~\cite{LHCb:2014acg,ALICE:2014eof,ALICE:2018oyo} a fit with two separate Pomerons is  practically excluded,  whereas the behavior  described by a single Pomeron  with a scale dependent intercept is valid up to the TeV region. A similar situation prevails for the photoproduction of $\Up$ mesons~\cite{LHCb:2015log}. In these two cases the hard scale is not introduced by the virtuality of the photon, but by the masses of the produced heavy quarks.

As long as we consider inclusive electroproduction processes with photon virtualities approximately above 3.5 GeV$^2$, as has been done in the HERA analysis~\cite{Adloff:2001rw, Radescu:2013mka}, it seems justified to identify the scale relevant for the Pomeron intercept with the photon virtuality, see {\it e.g.}, Ref.~\cite{Donnachie:2001wt}. The situation is more complex for exclusive diffractive reactions like photoproduction of vector mesons. There we have not only $Q^2=0$, but also the internal quark masses of the produced meson determine the scale. A relation between the scale of the intercept relevant for such exclusive processes and the scale relevant for inclusive processes has been derived in~\cite{Dosch:2015oha}; this relation is, however, not model-independent. For example, in Ref.~\cite{Dosch:2015oha} a connection was derived between the transverse size of the scattered object and the relevant scale for the Pomeron intercept.  Following this  connection, the intercept value $1.17$ was obtained for $J/\psi$ photoproduction and $1.25$ for $\Up$ production, in fair agreement with the data. This dependence of  the Pomeron parameters on the size of the scattered objects could also  explain the difference of the Pomeron intercept  between the proton and the pion which follows from the gluon distribution function (see Fig.~\ref{mudep}). One could indeed expect, that the effective scale for a pion is higher than that for a proton, since the pion is the smaller object.  At very high scales, corresponding to high virtuality  processes, this difference is supposed to  vanish, which  is supported by the theoretical results shown in Fig.~\ref{mudep}.

\begin{figure}
\includegraphics[width=10cm]{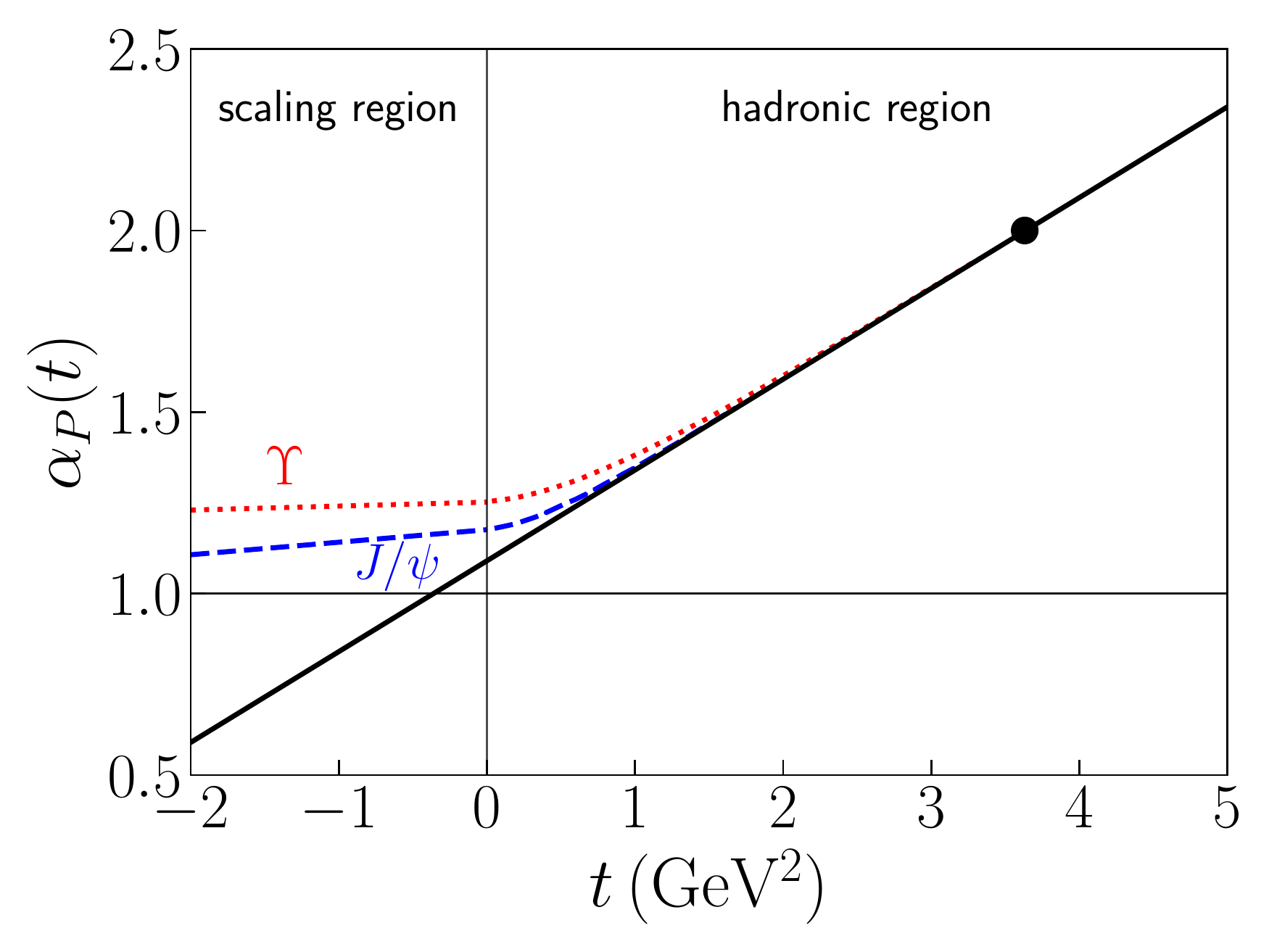}
\vspace{0cm}
\caption{\label{slope}  Scenario for Regge trajectories at different scales, from Ref.~\cite{Dosch:2015oha}. The solid line is the proposed trajectory for diffractive scattering of light hadrons or photoproduction of $\rho$ mesons; the dashed line is the trajectory for diffractive  photoproduction of $J/\psi$ mesons, and the dotted line that of $\Upsilon$ mesons. The scale is set by the transverse extension of the meson wave function. These trajectories are in agreement with experiments up to LHC energies.}
\end{figure}

In order to obtain information about the scale dependence of the Regge slope one has to  study the $t$-dependent generalized gluon distribution.  One can make, however, some general remarks.   Since the Regge trajectory for time-like $t$ values in the resonance region can be fixed by observable resonances, the linear slope in this region must be approximately independent on the scale. In order to have at $t=0$ a value  determined by the scale, the trajectory can therefore be nonlinear for all $t$ values. In Fig.~\ref{slope} a plausible scenario for Pomeron trajectories at different scales is presented. In the hadronic region ($t \gtrsim  1.5 \, {\rm GeV}^2$) the trajectory is fixed by its hadronic resonances, whereas in the scattering and production region ($t < 0$) it can depend on the scale, fixed, {\it e.g.,} by the transverse extension of the electroproduced object or the virtuality of the scattered photon.

%%%%%%%%%%%%%%%%%%%%%%%%%%%%%

%%%%%%%%%%%%%%%%%%%%%%%%%%%%

As can be seen from Fig.~\ref{slope}, the increase of the intercept can reduce the  trajectory slope for increasing scales. Such a decrease with increasing scale has been obtained qualitatively in the gauge/gravity dual model in Ref.~\cite{Brower:2006ea}.   It is also in accordance with the two-Pomeron approach~\cite{Donnachie:1998gm,Donnachie:2002en}, where the hard Pomeron has a  significantly smaller slope than the soft one.  In Ref.~\cite{Dosch:2015oha} the scale dependence of the Pomeron slope has been quantified. It has also been shown that at large space-like momentum transfer $t$, the trajectory $\alpha_P(t)$ of the Pomeron  approaches asymptotically a negative integer in order to analytically match the power-law behavior of the scattering amplitude at fixed $t/s$; {\it i.e.}, at fixed CM angles~\cite{Blankenbecler:1973kt}. 

We note that the full gravitational form factor is an observable quantity,  although the gluon and the quark components are individually scale dependent; only their sum is scale and renormalization scheme independent.  In Ref.~\cite{deTeramond:2021lxc}, the scale dependence of the gluon gravitational form factor is encoded in its  Fock state normalization, $c_\tau(\mu)$, which is equal to the gluon longitudinal momentum  using our normalization conventions. It is compensated by the scale dependence of the longitudinal momentum of the quarks by the momentum sum rule.

\section{\lb{SC}Summary and conclusion}

 We have studied the scale dependence $\mu$ of the Pomeron trajectory intercept $\alpha(0,\mu)$, which controls small-$x$ diffractive processes. To this end, we have related the Pomeron intercept to the scale evolution of the intrinsic gluon distribution function obtained in Ref.~\cite{deTeramond:2021lxc} in the framework of holographic light-from QCD,  together with the constraints imposed by the generalized Veneziano model. Our analysis assumes that the functional form of the  gluon distribution function is not modified by perturbative QCD evolution from the hadronic initial scale, where it is  normally  defined, to higher virtuality scales. This assumption is based on the observation that the application of  pQCD evolution to the intrinsic gluon contribution  indeed yields the full gluon distribution at all scales~\cite{deTeramond:2021lxc}; thus no additional perturbative Pomeron needs to be introduced. This assumption is also consistent with  the observed scale dependence of diffractive processes, since the  evolution of the gluon distribution leads to a scale-dependent Pomeron intercept.  This critical observation quantitatively explains the $Q^2$ dependence of the proton structure function $F_2(x, Q^2)$ with a single (unified) scale-dependent Pomeron exchange, and it  also  constitutes a basis for the  observed  scale dependence of diffractive electroproduction of vector mesons. Thus the  nonperturbative  ``soft" Pomeron with an intercept 1.08~\cite{Donnachie:1992ny} and the perturbative ``hard"  BFKL Pomeron~\cite{Fadin:1975cb,Kuraev:1977fs,Balitsky:1978ic} merge  into a single Pomeron with a scale-dependent intercept.  This scale dependence may seem unconventional, but it is perfectly compatible with the  foundations of Regge theory~\cite{Dosch:2015oha},  and it is conceptually satisfying.

 The results presented in this article lead to new insights into the essential scale dependence of the Pomeron trajectory underlying high energy, high virtuality processes, which in turn, provides a  unified framework for describing both the hard BFKL and soft Pomeron regimes. For the analysis of the Pomeron intercept, only the  gluon distribution function \req{g} at $t=0$ is relevant. The same procedure, can be extended, in principle, to study the scale dependence of the Pomeron slope, $\alpha'_P(\mu)$, by studying the pQCD evolution of the generalized gluon distribution function for non-vanishing momentum transfer $t$~\cite{deTeramond:2018ecg}.  In this case the full Pomeron trajectory $\alpha_P(t,\mu)$ enters, and we are confronted with the problem of  maintaining the scale invariance of the full gravitational  form factor at all $t$ values. This entails the study of delicate cancellations between the quark and gluon components, required to compute the scale dependence of the Pomeron slope $\alpha'_P(\mu)$ in terms of the  scale-dependence of the Pomeron intercept $\alpha(0,\mu).$  Further studies of photo-, and if possible electro-production processes at LHC energies would be very helpful in better understanding the nature of the Pomeron.

\acknowledgments{
T.L. is supported in part by National Natural Science Foundation of China under Contract No. 12175117. R.S.S. is supported  by U.S. DOE grant No. DE-FG02-04ER41302 and in part by the U.S. Department of Energy contract  No. DE-AC05-06OR23177,  under  which  Jefferson  Science  Associates, LLC, manages and operates Jefferson Lab. S.J.B. is supported in part by the Department of Energy Contract No. DE-AC02-76SF00515. A.D. is supported in part by the U.S. Department of Energy, Office of Science, Office of Nuclear Physics under Contract No. DE-AC05-06OR23177.
}

\end{document}